\documentclass[aps,prb,twocolumn,showpacs,groupedaddress,floatfix]{revtex4-1} 

\usepackage{amsmath}  
\usepackage{amsfonts} 
\usepackage{graphicx} 
\usepackage{latexsym}
\usepackage{amssymb}

\begin{document}


\title{A proper understanding of the Davisson and Germer experiments for undergraduate modern physics course}


\author{Masatsugu Suzuki}
\email{suzuki@binghamton.edu}
\affiliation{Department of Physics, State University of New York at Binghamton, Binghamton NY 13902-6000}

\author{Itsuko S. Suzuki}
\email{itsuko@binghamton.edu}
\affiliation{Department of Physics, State University of New York at Binghamton, Binghamton NY 13902-6000}


\date{\today}

\begin{abstract}
The physical interpretation for the Davisson-Germer experiments on nickel (Ni) single crystals [(111), (100), and (110) surfaces] is presented in terms of two-dimensional (2D) Bragg scattering. The Ni surface acts as a reflective diffraction grating when the incident electron beams hits the surface. The 2D Bragg reflection occurs when the Ewald sphere intersects the Bragg rods arising from the two-dimension character of the system. Such a concept is essential to proper understanding of the Davisson-Germer experiment for undergraduate modern physics course 
\end{abstract}

\maketitle 

\section{Introduction} 
\begin{figure}
\centering
\includegraphics[width=7.0cm]{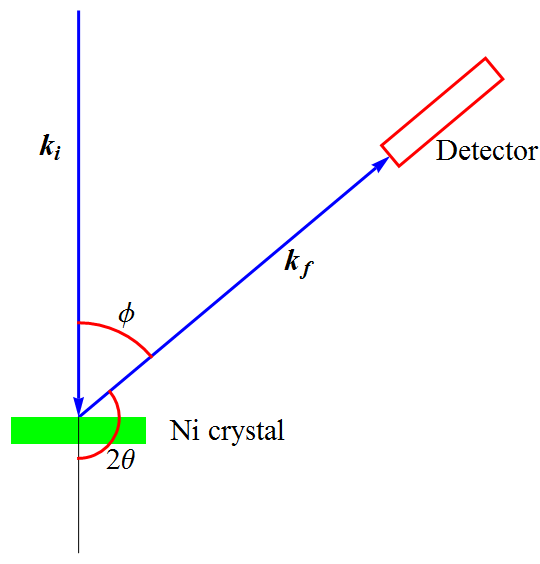}
\caption{Constructive interference of electron waves scattered from a single layer of Ni atoms (typically Ni (111) plane) at an angle $\phi$. $\mathbf{k}_{i}$ is wave vector of incident electron beam and $\mathbf{k}_{f}$ is wave vector of outgoing electron beam.}
\label{fig01}
\end{figure}

The observation of diffraction and interference of electron waves would provide the crucial test of the existence of wave nature of electrons. This observation was first seen in 1927 by C. J. Davisson and L. H. Germer.\cite{ref01} They studied electron scattering from a target consisting of a single crystal of nickel (Ni) and investigated this phenomenon extensively. Electrons from an electron gun are directed at a crystal and detected at some angle that can be varied (see Fig.\ref{fig01}). For a typical pattern observed, there is a strong scattering maximum at an angle of 50$^\circ$. The angle for maximum scattering of waves from a crystal depends on the wavelength of the waves and the spacing of the atoms in the crystal. Using the known spacing of atoms in their crystal, they calculated the wavelength that could produce such a maximum and found that it agreed with the de Broglie's hypothesis for the electron energy they were using. By varying the energy of the incident electrons, they could vary the electron wavelengths and produce maxima and minima at different locations in the diffraction patterns. In all cases, the measured wavelengths agreed with de Broglie's hypothesis.

The Davisson-Germer experiment itself is an established experiment.\cite{ref01,ref02,ref03,ref04,ref05,ref06} There is no controversy for them. How about the physical interpretation? One can see the description of the experiments and its physical interpretation in any standard textbook of the modern physics, which is one of the required classes for the physics majors (undergraduate) in U.S.A. Nevertheless, students as well as instructors in this course may have some difficulty in understanding the underlying physics, since the descriptions of the experiments are different depending on textbooks and are not always specific.\cite{ref07,ref08,ref09,ref10,ref11,ref12}

As far as we know, proper understanding has not been achieved fully so far. In some textbooks,\cite{ref09,ref10,ref12} the Ni layers are thought to act as a reflective diffraction grating. When electrons are scattered by the Ni (111) surface (single crystal), the electrons strongly interact with electrons inside the system. Thus electrons are scattered by a Ni single layer. The Ni (111) surface is just the two-dimensional layer for electrons. In other textbooks,\cite{ref07,ref08,ref11} electrons are scattered by Ni layers which act as a bulk system. The 3D character of the scattering of electrons appears in the form of Bragg points in the reciprocal lattice space.\cite{ref13,ref14,ref15,ref16,ref17,ref18} The 3D Bragg reflection can occur when the Bragg points lie on the surface of Ewald sphere, like the x-ray diffraction.

Here we will show that the Ni layers act as a reflective diffraction grating. The 2D scattering of electrons on the Ni (111), Ni(100), and Ni(110) surfaces will be discussed in terms of the concept of the Bragg rod (or Bragg ridge) which intersects the surcae of the Ewald sphere.\cite{ref13} We will show that the experimental results\cite{ref01,ref02,ref03,ref04,ref05} obtained by Davisson and Germer can be well explained in terms of this model. 

\section{\label{model}MODEL: EWALD SPHERE AND 2D BRAGG SCATTERING}

\begin{figure}
\centering
\includegraphics[width=7.0cm]{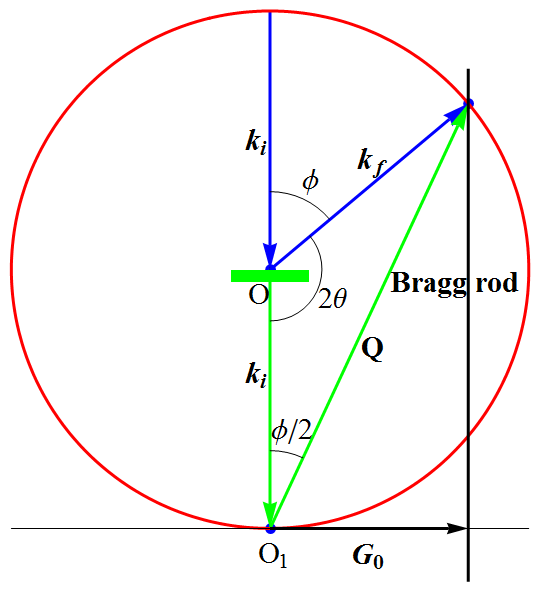}
\caption{Ewald sphere for the Bragg reflection for the 2D system. The wavevector $\mathbf{k}_{i}$ is drawn in the direction of the incident electron beam. $\mathbf{G}_{0}=\mathbf{G}_{\perp}$, which is the inplane reciprocal lattice vector, parallel to the surface. Ewald sphere (radius ($k_{f}=k_{i}=\frac{2\pi}{\lambda_{rel}}$) is centered at the point O. The point O$_{1}$ is the origin of the reciprocal lattice vectors. The Bragg reflection occurs when the surface of the Ewald sphere intersects the Bragg rod originated from the nature of the 2D system: $\phi=50^\circ$. $K = 54$ eV. $\lambda_{rel}=1.66891\AA$ for the Ni(111) plane. The lattice constant of conventional fcc Ni is $a=3.52\AA$.}
\label{fig02}
\end{figure}

\begin{figure}
\centering
\includegraphics[width=7.0cm]{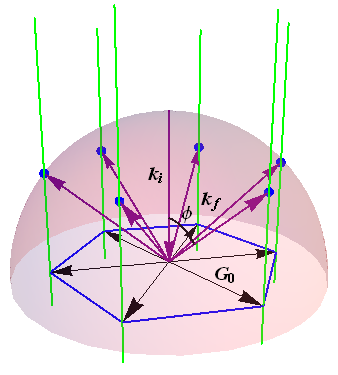}
\caption{Ewald sphere for the two-dimensional layer with the radius $k_{f}=k_{i}=\frac{2\pi}{\lambda_{rel}}$. The red lines are denoted by the Bragg rods arisen from the character of the 2D system. The Bragg reflection occurs when the wave vector of the reflected wave is on the point [denoted by the blue points, which are not the Bragg points], where the Ewald sphere intersects the Bragg rod. $\mathbf{G}_{0}=\mathbf{G}_{\perp}$.}
\label{fig03}
\end{figure}

In 1925, Davisson and Germer investigated the properties of Ni metallic surfaces by scattering electrons. Their experiments (Davisson-Germer experiment) demonstrates the validity of de Broglie's postulate because it can only be explained as a constructive interference of waves scattered by the periodic arrangement of the atoms of the crystal. The Bragg law for the diffraction had been applied to the x-ray diffraction, but this was first application to the electron waves.

We now consider the Bragg reflections in the 2D system. The Bragg reflections appear along the reciprocal rod, which is described by $\mathbf{G}_{\perp}$ , where $\mathbf{G}_{\perp}$ ($=\mathbf{G}_{0}$) is the in-plane reciprocal lattice vector parallel to the surface. The incident electron wave ($\mathbf{k}_{i}=\mathbf{k}$ , $k=2\pi/\lambda_{rel}$) is reflected by the surface of the 2D system. $\mathbf{k}_{f}$ ($=\mathbf{k}^\prime$) is the wavevector of the out-going electron wave ($k^\prime=2\pi/\lambda_{rel}$). Here we use the notation $\lambda_{rel}$ as the wavelength, instead of the conventional notation $\lambda$. The Ewald sphere is formed of the sphere with the radius of $k$ $(= k^\prime)$. The scattering vector $\mathbf{Q}$ is defined by
\begin{equation}
\mathbf{Q}=\mathbf{k}^\prime - \mathbf{k}  ,
    \label{eq01}
\end{equation}
and O$_{1}$ is the origin of the reciprocal lattice space. The 2D system is located at the origin of the real space O. The direction normal to the surface of the system is anti-parallel to the direction of the incident electron wave. Since the system is two-dimensional, the reciprocal lattice space is formed of Bragg rods. The Bragg reflections occur when the Bragg rods intersect the surface of the Ewald sphere.\cite{ref15,ref16}

Because of the 2D system, the Bragg points of the 3D system are changed into the Bragg rods. Then the Bragg condition occurs under the condition (see Fig.\ref{fig03}),
\begin{equation}
k^\prime \sin\phi=G_{\perp}=G_{0}  ,
    \label{eq02}
\end{equation}
where 
\begin{equation}
k=k^\prime =\frac{2\pi}{\lambda_{rel}}  .
    \label{eq03}
\end{equation}
The scattering angle $2\theta$ is related to the angle $\phi$ as
\begin{equation}
\phi=\pi-2\theta  .
    \label{eq04}
\end{equation}
In the electron diffraction experiment, we usually need to use the wavelength ( $\lambda_{rel}$), which is taken into account of the special relativity,\cite{ref07,ref08,ref09,ref10,ref11,ref12}
\begin{equation}
\sin(\pi-2\theta)=\frac{G_{0}}{k^\prime}=\frac{G_{0}}{2\pi}\lambda_{rel}  ,
    \label{eq05}
\end{equation}
or
\begin{equation}
\sin 2\theta=\frac{G_{0}}{2\pi}\lambda_{rel}  ,
    \label{eq06}
\end{equation}
where $\lambda_{rel}$ is the wavelength,
\begin{equation}
\lambda_{rel}=\frac{hc/E_{0}}{\sqrt{(K/E_{0})(K/E_{0}+2)}}  ,
    \label{eq07}
\end{equation}
where $h$ is the Planck's constant and $c$ is the velocity of light, $K$ (in the units of eV) is the kinetic energy of electron. $E_{0}$ ($= mc^{2}$) is the rest energy. $m$ is the rest mass of electron. In the non-relativistic limit, we have
\begin{equation}
\lambda_{classical}=\frac{12.2643}{\sqrt{K(\mbox{eV})}} \AA  ,
    \label{eq08}
\end{equation}
in the unit of $\AA$. When $K = 54$ eV, $\lambda_{rel}$ is calculated as $\lambda_{rel}= 1.66891\AA$.

Suppose that Ni (111) plane behaves like a three-dimensional system. The 3D Bragg reflection occurs only if the Bragg condition
\begin{equation}
\mathbf{Q}=\mathbf{k}_{f}-\mathbf{k}_{i}=\mathbf{G}  ,
    \label{eq09}
\end{equation}
is satisfied, where $\mathbf{Q}$ is the scattering vector and $\mathbf{G}$ is the reciprocal lattice vectors for the 3D system. In the experimental configuration as shown in Fig.\ref{fig02}. $\mathbf{G}$ is one of the reciprocal lattice vectors for the fcc Ni, and appears in the form of Bragg point. This Bragg point should be located on the surface of the Ewald sphere with radius ($k_{f}=k_{i}=2\pi/\lambda_{rel}$) centered at the point O (see Fig.\ref{fig02}). No existence of such a Bragg point on the Ewald sphere indicates that the 3D Bragg scattering does not occur in the present situation (Fig.\ref{fig02}).

\section{FUNDAMENTAL}
\subsection{Reciprocal lattice for the primitive cell for fcc}
The primitive cell by definition has only one lattice point. The primitive translation vectors of the fcc lattice are expressed by
\begin{equation}
\mathbf{a}_{1}=\frac{1}{2}a(0,1,1),
\mathbf{a}_{2}=\frac{1}{2}a(1,0,1),
\mathbf{a}_{3}=\frac{1}{2}a(1,1,0),
    \label{eq10}
\end{equation}
where there is one lattice point (or atom) per this primitive cell and $a$ is the lattice constant for the conventional cubic cell ($a = 3.52 \AA$ for fcc Ni).\cite{ref13} The corresponding reciprocal lattice vectors for the primitive cell are given by
\begin{subequations}
    \label{eq11}
\begin{eqnarray}
\mathbf{b}_{1}=\frac{2\pi(\mathbf{a}_{2}\times\mathbf{a}_{3})}{\mathbf{a}_{1}\cdot(\mathbf{a}_{2}\times\mathbf{a}_{3})}
=\frac{2\pi}{a}(-1,1,1)  ,\\
\mathbf{b}_{2}=\frac{2\pi(\mathbf{a}_{3}\times\mathbf{a}_{1})}{\mathbf{a}_{1}\cdot(\mathbf{a}_{2}\times\mathbf{a}_{3})}
=\frac{2\pi}{a}(1,-1,1)  ,\\
\mathbf{b}_{3}=\frac{2\pi(\mathbf{a}_{1}\times\mathbf{a}_{2})}{\mathbf{a}_{1}\cdot(\mathbf{a}_{2}\times\mathbf{a}_{3})}
=\frac{2\pi}{a}(1,1,-1)  .
\end{eqnarray}
The reciprocal lattice vector is described by
\begin{equation}
\mathbf{G}=g_{1}\mathbf{b}_{1}+g_{2}\mathbf{b}_{2}+g_{3}\mathbf{b}_{3}   ,
    \label{eq11d}
\end{equation}
\end{subequations}
where $g_{1}$, $g_{2}$, and $g_{3}$ are integers. 

\subsection{The reciprocal lattice for the conventional cell for fcc}
The translation vectors of the conventional unit cell (cubic) are expressed by
\begin{equation}
\mathbf{a}_{x}=a(1,0,0)  ,
\mathbf{a}_{y}=a(0,1,0)  ,
\mathbf{a}_{z}=a(0,0,1)  ,
    \label{eq12}
\end{equation}
where there are two atoms per this conventional unit cell.\cite{ref13} The reciprocal lattice vectors are defined by
\begin{subequations}
	\label{eq13}
\begin{eqnarray}
\mathbf{b}_{x}=\frac{2\pi(\mathbf{a}_{y}\times\mathbf{a}_{z})}{\mathbf{a}_{x}\cdot(\mathbf{a}_{y}\times\mathbf{a}_{z})}
=\frac{2\pi}{a}(1,0,0)  ,\\
\mathbf{b}_{y}=\frac{2\pi(\mathbf{a}_{z}\times\mathbf{a}_{x})}{\mathbf{a}_{x}\cdot(\mathbf{a}_{y}\times\mathbf{a}_{z})}
=\frac{2\pi}{a}(0,1,0)  ,\\
\mathbf{b}_{z}=\frac{2\pi(\mathbf{a}_{x}\times\mathbf{a}_{y})}{\mathbf{a}_{x}\cdot(\mathbf{a}_{y}\times\mathbf{a}_{z})}
=\frac{2\pi}{a}(0,0,1)  ,
\end{eqnarray}
In general, the reciprocal lattice vector is given by
\begin{equation}
\mathbf{G}=g_{x}\mathbf{b}_{x}+g_{y}\mathbf{b}_{y}+g_{z}\mathbf{b}_{z}
=\frac{2\pi}{a}(g_{x},g_{y},g_{z})  ,
	\label{eq13d}
\end{equation}
\end{subequations}
with
\begin{equation}
g_{x}=-g_{1}+g_{2}+g_{3}  , 
g_{x}=g_{1}-g_{2}+g_{3}  , 
g_{x}=g_{1}+g_{2}-g_{3}  . 
	\label{eq14}
\end{equation}
There are relations between $(g_{x}, g_{y}, g_{z})$ and $(g_{1}, g_{2}, g_{3})$. Note that all indices of $(g_{x}, g_{y}, g_{z})$ are odd or even. There is a selection rule for the indices $(g_{x}, g_{y}, g_{z})$.

\section{Structure factor for ideal 2D and 3D systems: Bragg rods and Bragg points}
The structure factor $S_{G}$ for the 2D system\cite{ref15,ref16} is given by
\begin{eqnarray}
S_{G}&=&\int n(\mathbf{r}_{2D})e^{-i\mathbf{G}\cdot \mathbf{r}_{2D}}d\mathbf{r}_{2D} \nonumber \\
&=&\int\int n(x,y)e^{-i(G_{x}x+G_{y}y)}dxdy  ,
	\label{eq15}
\end{eqnarray}
where
\[
\mathbf{r}_{2D}=x\mathbf{e}_{x}+y\mathbf{e}_{y}=(x,y,0)  .
\]
Then $S_{G}$ depends only on $G_{x}$ and $G_{y}$, forming the Bragg rod (or Bragg ridge) in the reciprocal lattice space. 

The structure factor $S_{G}$ for the 3D system\cite{ref13} is given by
\begin{eqnarray}
S_{G}&=&\int n(\mathbf{r}_{3D})e^{-i\mathbf{G}\cdot \mathbf{r}_{3D}}d\mathbf{r}_{3D}  \nonumber \\
&=&\int\int\int n(x,y,z)e^{-i(G_{x}x+G_{y}y+G_{z}z)}dxdydz  ,
	\label{eq16}
\end{eqnarray}
where $\mathbf{r}_{3D}$ is the position vectorof each atom,
\[
\mathbf{r}_{3D}=x\mathbf{e}_{x}+y\mathbf{e}_{y}+z\mathbf{e}_{z}=(x,y,z)  .
\]
Then $S_{\mathbf{G}}$ depends only on $G_{x}$, $G_{y}$, and $G_{z}$, which leads to the Bragg points.

Let $n_{j}(\mathbf{r}-\mathbf{r}_{j})$ be defined by the contribution of atom $j$ to the electron concentration. The electron concentration is expressed by
\begin{equation}
n(\mathbf{r})=\sum_{j=1}^{s}n_{j}(\mathbf{r}-\mathbf{r}_{j})  ,
	\label{eq17}
\end{equation}
over the $s$ atoms of the basis. Then we have
\begin{equation}
S_{\mathbf{G}}=\int_{V_{cell}}n(\mathbf{r})e^{-i\mathbf{G}\cdot \mathbf{r}}d\mathbf{r}
=\sum_{j}\int_{V_{cell}}n_{j}(\mathbf{r}-\mathbf{r}_{j})e^{-i\mathbf{G}\cdot \mathbf{r}}d\mathbf{r}  ,
	\label{eq18}
\end{equation}
or
\begin{equation}
S_{\mathbf{G}}=\sum_{j}e^{-i\mathbf{G}\cdot \mathbf{r}_{j}}\int_{V_{cell}}n_{j}\mathbf{(\rho)}e^{-i\mathbf{G}\cdot \mathbf{\rho}}d\mathbf{\rho}  .
	\label{eq19}
\end{equation}

\section{DISCUSSION}
\subsection{fcc Ni (111) plane}

\begin{figure}
\centering
\includegraphics[width=7.0cm]{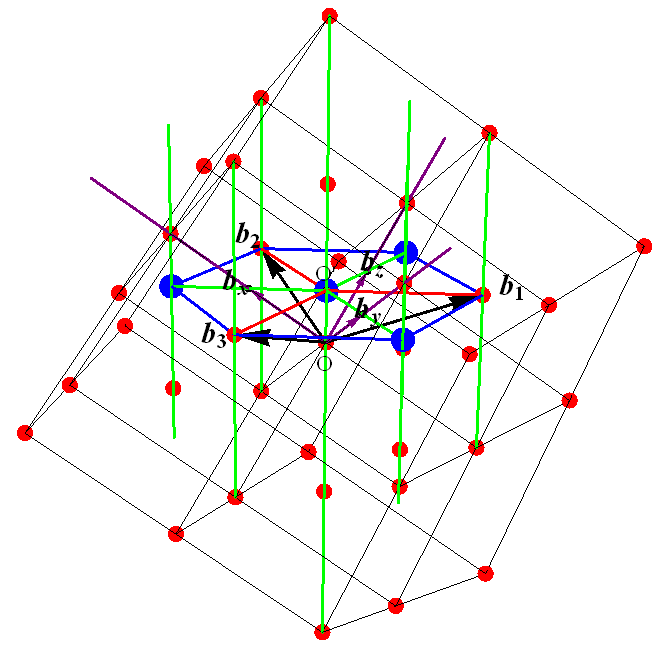}
\caption{The reciprocal lattice vectors which is viewed from the direction of ($\mathbf{b}_{1}+\mathbf{b}_{2}+\mathbf{b}_{3} =\mathbf{b}_{x} + \mathbf{b}_{y} + \mathbf{b}_{z}$) for Ni(111) plane. Note that $\mathbf{b}_{1}$, $\mathbf{b}_{2}$ and $\mathbf{b}_{3}$ are the reciprocal lattice vectors for the primitive cell where one atom exists, and $\mathbf{b}_{x}$, $\mathbf{b}_{y}$, and $\mathbf{b}_{z}$ are the reciprocal lattice vectors for the conventional cell. 2D Reciprocal lattice plane, which is viewed from the direction of ($\mathbf{b}_{1}+\mathbf{b}_{2}+\mathbf{b}_{3} = \mathbf{b}_{x} +\mathbf{b}_{y} + \mathbf{b}_{z}$). The green lines form a Bragg rod along the direction of ($\mathbf{b}_{1}+\mathbf{b}_{2}+\mathbf{b}_{3} = \mathbf{b}_{x} + \mathbf{b}_{y} + \mathbf{b}_{z}$), arising from the 2D character of the system. The red circle shows the 3D Bragg point of fcc Ni. The blue circle is not the 3D Bragg point and lies on the 2D Bragg rods. }
\label{fig04}
\end{figure}

\begin{figure}
\centering
\includegraphics[width=7.0cm]{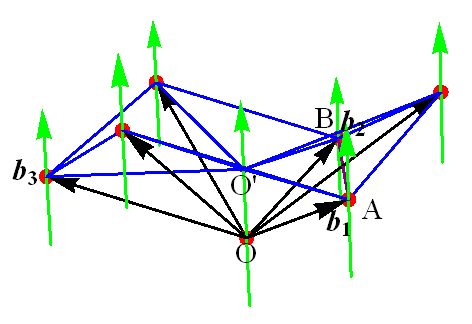}
\caption{2D reciprocal lattice plane formed by Bragg rods, where the green arrows are the Bragg rod along the from the direction of ($\mathbf{b}+\mathbf{b}_{2}+\mathbf{b}_{3} = \mathbf{b}_{x} + \mathbf{b}_{y} + \mathbf{b}_{z}$) [Ni(111) plane]. Bragg rod forming along the direction ($\mathbf{b}_{1}+\mathbf{b}_{2}+\mathbf{b}_{3} = \mathbf{b}_{x} + \mathbf{b}_{y} + \mathbf{b}_{z}$). The red circle denotes the 3D Bragg point of fcc Ni. }
\label{fig05}
\end{figure}

Here we discuss the experimental results obtained by Davisson and Germer in terms of the model described in the Section \ref{model}.

\begin{figure}
\centering
\includegraphics[width=7.0cm]{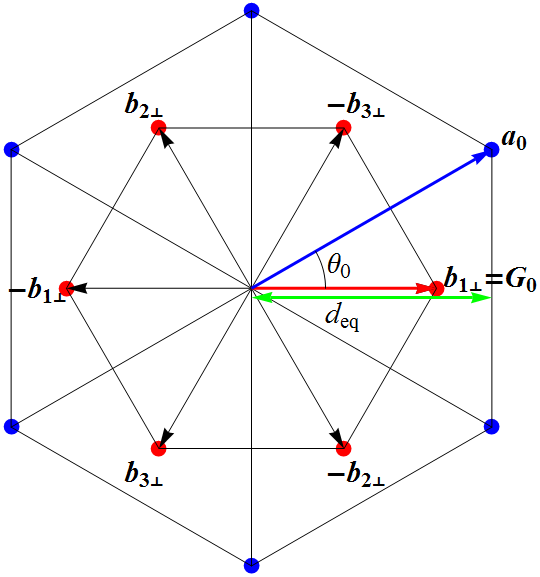}
\caption{The 2D reciprocal lattice vector formed by Bragg rods in the case of corresponding to the Ni (111) plane. The corresponding 2D lattice vectors in the real space are also shown. The axis of $\mathbf{a}_{0}$ is rotated by 30$^\circ$ with respect to the axis of the reciprocal lattice $\mathbf{G}_{0}\cdot\theta_{0} = 30^\circ$. $\mathbf{G}_{0} \cdot \mathbf{a}_{0} =2\pi$. $G_{0}=\frac{4\pi}{\sqrt{3}a_{0}} = 2.915\AA^{-1}$. $a_{0}=\frac{a}{\sqrt{2}} =2.489\AA$. $d_{eq}=\frac{a_{0}}{\sqrt{2}}=a\sqrt{\frac{3}{8}}$. $d_{eq}$ ($= 2.1556\AA$) is the distance such that Ni (111) plane acts as a reflective diffraction grating,\cite{ref02} $d_{eq} \sin \phi =\lambda _{rel}$.}
\label{fig06}
\end{figure}

\begin{figure}
\centering
\includegraphics[width=8.0cm]{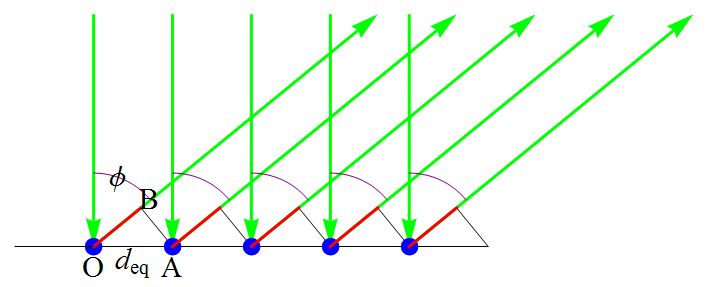}
\caption{Reflective diffraction grating. $d_{eq}(111)=2.15\AA$ for Ni(111) plane.\cite{ref02} $\phi = 50.74^\circ$. $a=3.52\AA$ for Ni. The blue points denote Ni atoms on the 2D layer.}
\label{fig07}
\end{figure}

Here we note that
\begin{equation}
\mathbf{b}_{1}=\frac{2\pi}{a}(-1,1,1)  ,
\mathbf{b}_{2}=\frac{2\pi}{a}(1,-1,1)  ,
\mathbf{b}_{3}=\frac{2\pi}{a}(1,1,-1)  ,
	\label{eq20}
\end{equation}
with
\begin{equation}
\mathbf{b}_{1}+\mathbf{b}_{2}+\mathbf{b}_{3}=\frac{2\pi}{a}(1,1,1)  .
	\label{eq21}
\end{equation}
The unit vector along the direction of the vector $\overrightarrow{OO^\prime}$ is given by
\begin{equation}
\hat{\mathbf{n}}(111)=\frac{\overrightarrow{OO^\prime}}{\overline{OO^\prime}}=\frac{1}{\sqrt{3}}(1,1,1)  .
	\label{eq22}
\end{equation}
The component of $\mathbf{b}_{1}$ parallel to the unit vector $\hat{\mathbf{n}}(111)$ is
\begin{equation}
\mathbf{b}_{1\parallel}=\lbrack \hat{\mathbf{n}}(111)\cdot \mathbf{b}_{1} \rbrack \hat{\mathbf{n}}(111)
=\frac{2\pi}{3a}(1,1,1)  .
	\label{eq24}
\end{equation}
Similarly, we have
\begin{equation}
\mathbf{b}_{2\parallel}=\mathbf{b}_{3\parallel}=\frac{2\pi}{3a}(1,1,1)  ,
\end{equation}
which is equal to 
\begin{equation}
\overrightarrow{OO^\prime}=\frac{\mathbf{b}_{1}+\mathbf{b}_{2}+\mathbf{b}_{3}}{3} .
	\label{eq25}
\end{equation}
The component of $\mathbf{b}_{1}$, $\mathbf{b}_{2}$, and $\mathbf{b}_{3}$, perpendicular to the unit vector $\hat{\mathbf{n}}(111)$ are
\begin{subequations}
	\label{eq26}
\begin{eqnarray}
\mathbf{b}_{1\perp}=\mathbf{b}_{1}-\mathbf{b}_{1\parallel}=\frac{4\pi}{3a}(-2,1,1)  ,	\label{eq26a} \\
\mathbf{b}_{2\perp}=\mathbf{b}_{2}-\mathbf{b}_{2\parallel}=\frac{4\pi}{3a}(1,-2,1)  ,	\label{eq26b} \\
\mathbf{b}_{3\perp}=\mathbf{b}_{3}-\mathbf{b}_{3\parallel}=\frac{4\pi}{3a}(1,1,-2)  .	\label{eq26c} 
\end{eqnarray}
\end{subequations}
Then we get
\begin{subequations}
	\label{eq28}
\begin{eqnarray}
\mathbf{b}_{1\perp}+\mathbf{b}_{2\perp}=-\mathbf{b}_{3\perp }=\frac{4\pi}{3a}(-1,-1,2) ,  \label{eq28a}\\ 
\mathbf{b}_{2\perp}+\mathbf{b}_{3\perp}=-\mathbf{b}_{1\perp }=\frac{4\pi}{3a}(2,-1,-1) , \label{eq28b}\\ 
\mathbf{b}_{3\perp}+\mathbf{b}_{1\perp}=-\mathbf{b}_{2\perp }=\frac{4\pi}{3a}(-1,2,-1) . \label{eq28c} 
\end{eqnarray}
\end{subequations}
The 2D reciprocal lattice vector formed by Bragg rods ($\mathbf{b}_{1\perp}$, $\mathbf{b}_{2\perp}$, $\mathbf{b}_{3\perp}$, $-\mathbf{b}_{1\perp}$, $-\mathbf{b}_{2\perp}$, $-\mathbf{b}_{3\perp}$) is shown by Figs.\ref{fig04} and \ref{fig05}, where the magnitude of the reciprocal lattice vector is given by
\begin{equation} 
\left| \overrightarrow{O^\prime A} \right| =G_{0}=\left| b_{1\perp} \right|
=\frac{4\pi }{\sqrt{3} a_{0} } ,
	\label{eq29}
\end{equation}
where $a_{0}=a/\sqrt{2}$.
Note that $\mathbf{G}_{0}$ can be also obtained as
\begin{equation} 
\mathbf{G}_{0}=\mathbf{b}_{1}-\frac{1}{3}(\mathbf{b}_{1}+\mathbf{b}_{2}+\mathbf{b}_{3})=\frac{4\pi }{3a}(-2,1,1)  .
	\label{eq31}
\end{equation}
Figure \ref{fig06} shows the 2D reciprocal lattice vectors formed by the Bragg rods with the six-fold symmetry. This implies that the corresponding 2D triangular lattice is formed in the real space. The direction of the fundamental lattice vector $\mathbf{a}_{0}$ is rotated by 30$^\circ$ with respect to the direction of the fundamental reciprocal lattice vector $\mathbf{G}_{0}$,\cite{ref13} where
\begin{equation} 
\mathbf{a}_{0} \cdot \mathbf{G}_{0} =2\pi .
	\label{eq32}
\end{equation}
Using the geometry as shown in Fig.\ref{fig03}, the Bragg condition can be obtained as
\begin{equation} 
\sin (2\theta )=\sin \phi =nG_{0} \frac{\lambda _{rel} }{2\pi } 
=n\frac{\lambda _{rel} }{a} \sqrt{\frac{8}{3} } ,
	\label{eq33}
\end{equation}
where $n$ = 1, 2, , 3,..... and $G_{0}$ is the fundamental reciprocal lattice (see Fig.\ref{fig06}). Note that $n = \sqrt{3}$ and $2\sqrt{3}$ are also possible for $\sqrt{3}G_{0}$ and $2\sqrt{3}G_{0}$, respectively. Here we only consider the case of integer $n$. We introduce the length $d_{eq}(111)$ such that
\begin{equation} 
d_{eq}(111) \sin \phi =n\lambda _{rel} ,
	\label{eq34}
\end{equation}
where
\begin{equation} 
d_{eq}(111) =a\sqrt{\frac{3}{8} } =\frac{\sqrt{6}a}{4} = 0.6124 \times 3.52 \AA=2.1556 \AA  .
	\label{eq35}
\end{equation}
Equation (\ref{eq34}) with $n=1$ corresponds to the expression for the reflective diffraction grating, where 
\begin{equation} 
d_{eq}(111) \sin \phi =\lambda _{rel} ,
	\label{eq36}
\end{equation} 
for the Ni(111) plane. This value of $d_{eq}$ agrees well with that reported by Davisson and Germer.\cite{ref01,ref02} We note that the left side of Eq.(\ref{eq36}) is the path difference between two adjacent rays for the reflective diffraction grating (see Fig.\ref{fig07}). When $K$ = 54 eV, the wavelength can be calculated as $\lambda_{rel} = 1.6689 \AA$, using Eq.(\ref{eq07}). From the result of the Davisson-Germer experiment,\cite{ref01,ref02} $\phi=50.74^\circ$. we get $\lambda _{exp } =d_{eq}(111) \sin (\phi )= 1.6684 \AA$. This wavelength is exactly the same as that calculated based on the de Broglie hypothesis. 

\begin{figure}
\centering
\includegraphics[width=8.0cm]{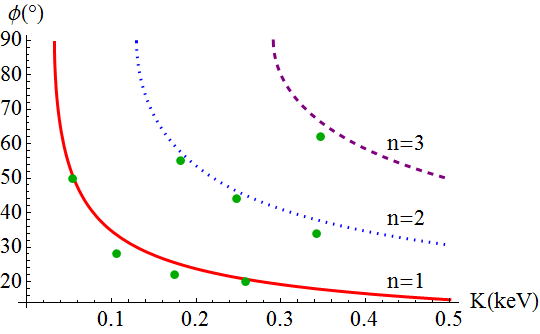}
\caption{The angle $\phi$ vs the kinetic energy $K$ for the Ni (111) plane. The data denoted by points (green) were reported by Davisson and Germer.\cite{ref01,ref02} The red solid line for $\mathbf{G}_{0}$ ($n=1$). The blue dotted line for $2\mathbf{G}_{0}$ ($n=2$). The purple dashed line for $3 \mathbf{G}_{0}$ ($n=3$), where $G_{0} =\frac{4\pi }{\sqrt{3}a}\sqrt{2}=2.9149\AA^{-1}$.}
\label{fig08}
\end{figure}
 
Figure \ref{fig08} shows the plot of the angle $\phi$ as a function of the kinetic energy $K$, which is expressed by Eq.(\ref{eq33}), where $n$ = 1, 2, and 3. In Fig.\ref{fig08}, we also plot the experimental data obtained by Davisson and Germer (denoted by green points). We find that all the data lie well on the predicted relation between $\phi$ and $K$ for $n$ = 1, 2, and 3.
 
The six-fold symmetry of the 2D reciprocal lattice vectors was experimentally confirmed by Davisson and Germer for the Ni(111) plane [$K$ = 54 eV and $\phi=50^\circ$].\cite{ref01,ref02} The rotation of the Ni sheet around the (111) direction leads to nealy six-fold symmetry of the intensity as a function of azimuthal for latitude. Note that the intensities at $\mathbf{Q}_{\perp}=\mathbf{b}_{1\perp}$, $\mathbf{b}_{2\perp }$, and $\mathbf{b}_{3\perp}$ (denoted as (111) plane by Davissson and Germer)\cite{ref01} are stronger than those from $\mathbf{Q}_{\perp}=(\mathbf{b}_{1} +\mathbf{b}_{2} )_{\perp}$, $(\mathbf{b}_{2}+\mathbf{b}_{3})_{\perp}$, and $(\mathbf{b}_{3}+\mathbf{b}_{1})_{\perp}$ (denoted as (100) plane by Davisson and Germer).\cite{ref01} We also note that when $K$ = 65 eV, Davisson and Germer observed $\phi=44^\circ$, where $\lambda_{rel}$ can be evaluated as $\lambda_{rel}=1.5212\AA$, using Eq.(\ref{eq36}) with $d_{eq}(111)$ by Eq.(\ref{eq35}).\cite{ref01} In this case, the intensities at $\mathbf{Q}_{\perp}=\mathbf{b}_{1\perp}$, $\mathbf{b}_{2\perp}$, and $\mathbf{b}_{3\perp}$ are much weaker than those from $\mathbf{Q}_{\perp}=(\mathbf{b}_{1} +\mathbf{b}_{2})_{\perp}$, $(\mathbf{b}_{2}+\mathbf{b}_{3})_{\perp}$, and $(\mathbf{b}_{3} +\mathbf{b}_{1})_{\perp}$. In other words, the intensity vs azimuthal pattern is strongly dependent of the kinetic energy of electrons. For the ideal case of scattering from a true 2D network of atoms, the intensity vs azimuthal should show the perfect six-fold symmetry. The intensity is the same for $\mathbf{Q}_{\perp}=\mathbf{b}_{1\perp}$, $\mathbf{b}_{2\perp}$, $\mathbf{b}_{3\perp}$, $(\mathbf{b}_{1}+\mathbf{b}_{2})_{\perp}$, $(\mathbf{b}_{2} +\mathbf{b}_{3})_{\perp}$, and $(\mathbf{b}_{3}+\mathbf{b}_{1})_{\perp}$. In the Davisson-Germer experiment,\cite{ref01} it may be possible that the primary electrons penetrate several atomic layers into the system. The deeper they penetrate, the more scattering events in the direction perpendicular to the surface, enhancing the contribution of the 3D scattering to experimental results. This leads to the change of the intensity of the Bragg reflections as a function of azimuthal, in comparison with the case of pure 2D scattering.\cite{ref16} 

\subsection{Ni (100) plane}

\begin{figure}
\centering
\includegraphics[width=7.0cm]{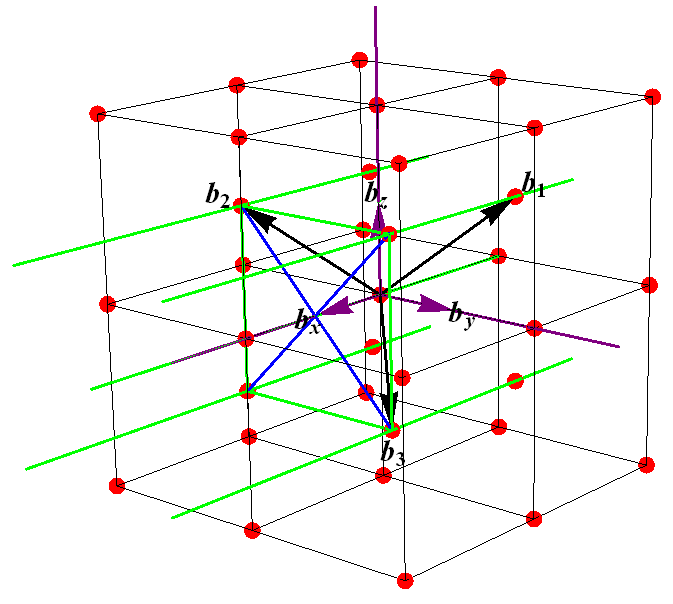}
\caption{Reciprocal lattice plane which is viewed from the $\mathbf{b}_{x}$-direction, where $\mathbf{b}_{x}$ is the reciprocal lattice vector of the conventional cubic lattice. Ni(100) plane. The red circle denotes the 3D Bragg points for fcc Ni.}
\label{fig09}
\end{figure}

\begin{figure}
\centering
\includegraphics[width=7.0cm]{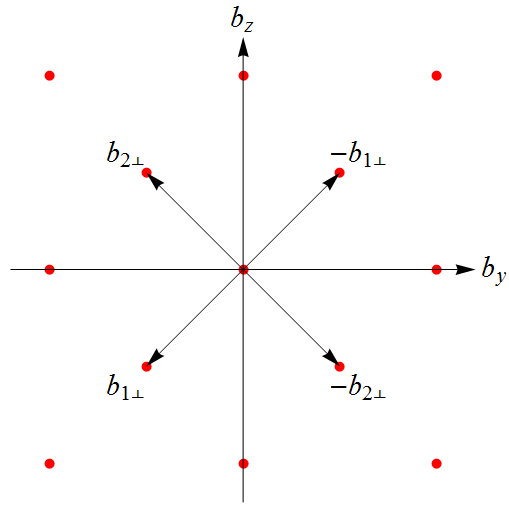}
\caption{The 2D reciprocal lattice vector formed by Bragg rods in the case of corresponding to the Ni (100) plane. $\mathbf{b}_{1\perp}$, $\mathbf{b}_{2\perp}$ are the reciprocal lattice vectors, which is viewed from the $\mathbf{b}_{x}$-direction, where $\mathbf{b}_{x}$ is the reciprocal lattice vector of the conventional cell. $\hat{\mathbf{n}}(100)=(1,0,0)$. }
\label{fig10}
\end{figure}
 
The unit vector along the (1,0,0) direction is defined by
\begin{equation}
\hat{\mathbf{n}}(100)=(1,0,0)  .
	\label{eq39}
\end{equation}
The components of $\mathbf{b}_{1}$ and $\mathbf{b}_{2}$, parallel to the unit vector $\hat{\mathbf{n}}(100)$ are
\begin{subequations}
	\label{eq40}
\begin{eqnarray} 
\mathbf{b}_{1\parallel}=\lbrack\hat{\mathbf{n}}(100)\cdot\mathbf{b}_{1}\rbrack\hat{\mathbf{n}}(100)
=\frac{2\pi}{a}(-1,0,0) ,  \label{eq40a}\\
\mathbf{b}_{2\parallel}=\lbrack\hat{\mathbf{n}}(100)\cdot\mathbf{b}_{2}\rbrack\hat{\mathbf{n}}(100)
=\frac{2\pi}{a}(1,0,0) .  \label{eq40b}
\end{eqnarray}
Then the components of $\mathbf{b}_{1}$ and  $\mathbf{b}_{2}$ perpendicular to the unit vector $\hat{\mathbf{n}}(100)$ are
\begin{eqnarray}
\mathbf{b}_{1\perp}=\mathbf{b}_{1}-\mathbf{b}_{1\parallel}=\frac{2\pi}{a}(0,-1,-1)  , \label{eq40c}\\
\mathbf{b}_{2\perp}=\mathbf{b}_{2}-\mathbf{b}_{2\parallel}=\frac{2\pi}{a}(0,-1,1)  . \label{eq40d}
\end{eqnarray}
\end{subequations} 
Then we get the 2D reciprocal lattice vectors formed by Bragg rods, having the four-fold symmetry around the vector $\hat{\mathbf{n}}(100)$, 
\begin{equation} 
\left|\mathbf{b}_{1\perp}\right|=\left|\mathbf{b}_{2\perp}\right|=G_{0}=\frac{2\pi}{a}\sqrt{2} .
	\label{eq41}
\end{equation}
Using the geometry as shown in Fig.\ref{fig10}, the Bragg condition can be expressed in terms of
\begin{equation} 
\sin (2\theta )=\sin \phi =nG_{0} \frac{\lambda _{rel} }{2\pi } 
=\frac{n\lambda _{rel} }{d_{eq} } ,\
	\label{eq42}
\end{equation} 
for the Ni(100) plane, where $d_{eq}(100)$ is the length of spacing for the reflective diffraction grating for Ni(100) plane $d_{eq}(100) =\frac{a}{\sqrt{2}}= 2.517\AA$.

\begin{figure}
\centering
\includegraphics[width=8.0cm]{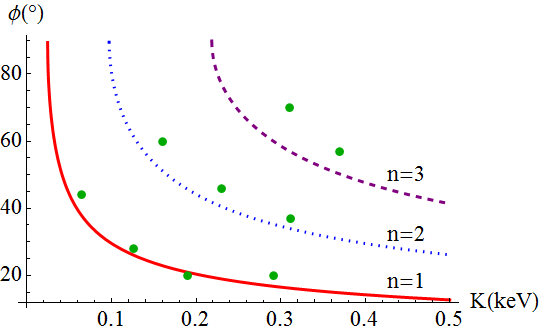}
\caption{The angle $\phi$ vs the kinetic energy $K$ for the Ni (100) plane. The data denoted by points (green) were reported by Davisson and Germer.\cite{ref01,ref02} The red solid line for $\mathbf{G}_{0}$. The blue dotted line for $2\mathbf{G}_{0}$. The purple dashed line for $3\mathbf{G}_{0}$, where $G_{0}=\frac{2\pi}{a}\sqrt{2}=2.5244\AA^{-1}$.}
\label{fig11}
\end{figure}

Figure \ref{fig11} shows the plot of the angle $\phi$ as a function of the kinetic energy $K$, which is expressed by Eq.(\ref{eq42}), where $n$ = 1, 2, and 3. In Fig.\ref{fig11}, we also plot the experimental data obtained by Davisson and Germer (denoted by green points).\cite{ref01,ref02} We find that all the data fall fairly well on the predicted relation between $\phi$ and $K$ for $n$ = 1, 2, and 3, in particular for $n=1$. When $K$ = 190 eV, the wavelength can be calculated as $\lambda_{rel} = 0.88966 \AA  $, using Eq.(\ref{eq07}). From the result of the Davisson-Germer experiment, $\phi = 20^\circ$,\cite{ref01,ref02} on the other hand, we get $\lambda_{exp}=d_{eq}(100)\sin\phi=0.86087 \AA$ for the Ni(100) plane. This wavelength is almost the same as that calculated based on the de Broglie's hypothesis.

\subsection{Ni (110) plane}

\begin{figure}[h!]
\centering
\includegraphics[width=7.0cm]{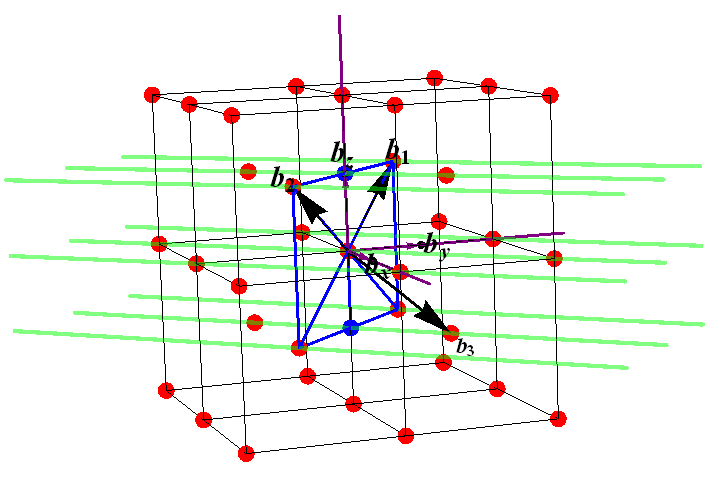}
\caption{Reciprocal lattice plane which is viewed from the $\mathbf{b}_{x}$ + $\mathbf{b}_{y}$ direction, where $\mathbf{b}_{x}$ and $\mathbf{b}_{y}$ are the reciprocal lattice vector of the conventional fcc lattice. Ni(110) plane. The red circle denotes the 3D Bragg point. The blue circle does not denote the 3D Bragg point and lies on the 2D Bragg rod. }
\label{fig12}
\end{figure}
 
The unit vector along the (110) direction is defined by
\begin{equation} 
\hat{\mathbf{n}}(110)=\frac{1}{\sqrt{2}}(1,1,0)  .
	\label{ad44}
\end{equation}
The components of $\mathbf{b}_{1}$, $\mathbf{b}_{1}$, and $\mathbf{b}_{1}$, parallel to the unit vector $\hat{\mathbf{n}}(110)$ are
\begin{subequations}
	\label{ad45}
\begin{eqnarray}
\mathbf{b}_{1\parallel}=\lbrack\hat{\mathbf{n}}(110)\cdot\mathbf{b}_{1}\rbrack\hat{\mathbf{n}}(110)
=(0,0,0)  ,	\label{ad45a}\\
\mathbf{b}_{2\parallel}=\lbrack\hat{\mathbf{n}}(110)\cdot\mathbf{b}_{2}\rbrack\hat{\mathbf{n}}(110)
=(0,0,0)  ,	\label{ad45b}\\
\mathbf{b}_{3\parallel}=\lbrack\hat{\mathbf{n}}(110)\cdot\mathbf{b}_{3}\rbrack\hat{\mathbf{n}}(110)
=\frac{2\pi}{a}(1,1,0)  .	\label{ad45c}
\end{eqnarray}
The components of $\mathbf{b}_{1}$, $\mathbf{b}_{2}$, and $\mathbf{b}_{3}$, perpendicular to the unit vector $\hat{\mathbf{n}}(110)$ are
\begin{eqnarray}
\mathbf{b}_{1\perp}=\mathbf{b}_{1}-\mathbf{b}_{1\parallel}=\frac{2\pi}{a}(-1,1,1)  ,  \label{ad45d} \\
\mathbf{b}_{2\perp}=\mathbf{b}_{2}-\mathbf{b}_{2\parallel}=\frac{2\pi}{a}(1,-1,1)  ,  \label{ad45e} \\
\mathbf{b}_{3\perp}=\mathbf{b}_{3}-\mathbf{b}_{3\parallel}=\frac{2\pi}{a}(0,0,-1)  .  \label{ad45f} 
\end{eqnarray}
\end{subequations} 
Then we get the magnitude of the 2D reciprocal lattice vector (rectangular lattice)
\begin{equation} 
G_{0} =\frac{\left| \mathbf{b}_{1\perp}-\mathbf{b}_{2\perp}\right|}{2} =\frac{2\pi}{a} \sqrt{2} ,
G_{1} =\left| \mathbf{b}_{3\perp} \right| =\frac{2\pi}{a} .
	\label{eq46}
\end{equation}

\begin{figure}
\centering
\includegraphics[width=8.0cm]{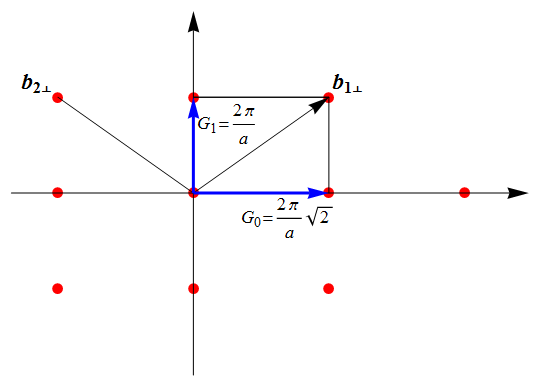}
\caption{The 2D reciprocal lattice vector formed by Bragg rods in the case of corresponding to the Ni (110) plane. $\mathbf{b}_{1\perp}$, $\mathbf{b}_{2\perp}$ are the reciprocal lattice vectors, which is viewed from the $\hat{\mathbf{n}}(110)=\frac{1}{\sqrt{2}}(1,1,0)$ direction.}
\label{fig13}
\end{figure}
 
\begin{figure}
\centering
\includegraphics[width=8.0cm]{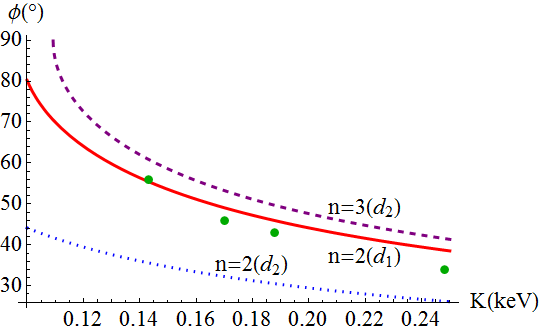}
\caption{The angle $\phi$ vs the kinetic energy $K$ for the Ni (110) plane. The data denoted by points (green) were reported by Davisson and Germer.\cite{ref01,ref02} $\sin \phi =n\frac{\lambda_{rel}}{d_{1} } $ [Eq.(\ref{eq47})] with $n = 2$, red solid line). $\sin \phi =n\frac{\lambda_{rel}}{d_{2}}$ ($n = 2$, blue dotted line). $\sin \phi =n\frac{\lambda_{rel}}{d_{2}}$ ($n = 3$, purple dashed line).}
\label{fig14}
\end{figure}

Using the geometry as shown in Fig.\ref{fig13}, the Bragg conditions for $n\mathbf{G}_{0}$ and $n\mathbf{G}_{1}$ can be expressed by
\begin{equation} 
\sin (2\theta )=\sin \phi 
=nG_{0} \frac{\lambda_{rel}}{2\pi} 
=n\frac{\lambda_{rel}}{a} \sqrt{2} =\frac{n\lambda_{rel}}{d_{1}}  ,
	\label{eq47}
\end{equation} 
and
\begin{equation} 
\sin (2\theta )=\sin \phi 
=nG_{1} \frac{\lambda _{rel} }{2\pi } 
=n\frac{\lambda_{rel}}{a} =\frac{n\lambda_{rel}}{d_{2} } ,
	\label{eq48}
\end{equation} 
respectively, where $d_{1} =\frac{a}{\sqrt{2}}=2.517\AA$ and $d_{2}=a=3.52\AA$.
The lengths $d_{0}$ and $d_{1}$ are equivalent spacings of the 2D rectangular lattice (real space). Figure \ref{fig14} shows the plot of the angle $\phi$ as a function of the kinetic energy $K$, which is expressed by Eq.(\ref{eq47}), where $n$ = 1, 2, and 3. In Fig.\ref{fig14}, we also plot the experimental data obtained by Davisson and Germer (denoted by green points).\cite{ref01,ref02} We find that all the data lie fairly well on the predicted relation given by Eq.(\ref{eq47}) between $\phi$ and $K$ for $\sin \phi =n\frac{\lambda_{rel}}{d_{1}}$ with $n$ = 2. 

When $K$ = 143 eV, the wavelength can be calculated as $\lambda_{rel} = 1.0255 \AA$, using Eq.(\ref{eq07}). From the result of the Davisson-Germer experiment, $\phi = 56^\circ$, on the other side, we get
\begin{equation} 
\lambda_{exp}=\frac{d_{1}}{2}\sin\phi=\frac{a}{2\sqrt{2}}\sin\phi= 1.0317 \AA ,
	\label{eq51}
\end{equation} 
using $a= 3.52\AA$. This wavelength is almost the same as that calculated based on the de Broglie's hypothesis. We note that the $d$-spacing $d_{eq}(110)$ for the reflective diffraction grating is $d_{eq}(110) =\frac{a}{2\sqrt{2}} =1.2445 \AA$, for the Ni(110) plane. This value of $d_{eq}(110)$ agrees well with that reported by Davisson and Germer.\cite{ref02} 

\section{CONCLUSION}
The essential feature of the Davisson-Germer experiment for the Ni(111), Ni(100), and Ni(110) planes is that the 2D Bragg scattering occurs. The Bragg rods are formed in the reciprocal lattice space. The component of the scattering vector \textbf{\textit{Q}} parallel to the surface is equal to the 2D surface reciprocal lattice vector of the Bragg rods. The electron beam is reflected from a single layer, leading to the eflective diffraction grating with the $d$-spacing $d_{eq}$.

\end{document}